\documentclass[showpacs,longbibliography]{svproc}
\usepackage[utf8]{inputenc}

\usepackage{graphicx}
\usepackage{multicol}
\usepackage{footmisc}
\usepackage[english]{babel}
\usepackage{float}
\usepackage{amsmath,amssymb,enumerate,bm}
\usepackage{url,bbding}

\begin{document}
\mainmatter

\title{Statistical properties of a granular gas fluidized by turbulent air wakes}
\titlerunning{Air fluidized granular gas}

\author{Miguel A. López-Castaño \Envelope \and Juan F. González-Saavedra \and \'Alvaro
  Rodríguez-Rivas \and Francisco Vega Reyes}
\authorrunning{López-Castaño \textit{et al.}}

\institute{Departamento de Física and Instituto de Computación Científica Avanzada (ICCAEx), Universidad de Extremadura, 06071 Badajoz, Spain \\
\email{malopez00@unex.es}}
\date{\today}

\maketitle

\begin{abstract}
    We perform experiments with a granular system that
    consists of a collection of identical hollow spheres (ping-pong balls). 
    Particles rest on a horizontal metallic grid and are confined within a
    circular region. Fluidization is achieved by means of a turbulent air current
    coming from below. Air flow is adjusted so that the balls do not
    elevate over the grid, as an approach to 2D dynamics. With a high-speed camera, we take images of the system. From these images we can infer horizontal particle positions and velocities by means of particle-tracking algorithms. 
    With the  obtained data we analyze: a) the
    systematic measurement error in the determination of positions and
    velocities from our digital images; b) the degree of homogeneity
    achieved in our experiments (which depends on possible deviations of the grid from the
    horizontal and on the homogeneity of turbulent air wakes). Interestingly, we have observed evidences of crystallization at high enough densities.
\end{abstract}

\section{Introduction}

Experimental works on the symmetry properties of nearly close-packed particles
were carried out in the early 20th for direct visualization of the crystal structure in laboratory scale model
systems (by that time this was technically not possible for real solid
crystals). Pioneering work by L. Bragg and J. F. Nye in experiments with soap bubbles monolayers \cite{BN47},
which clearly reported a macroscopic hexagonal crystal structure, was followed by other
interesting analogous works on different phases in systems with macroscopic particles \cite{F64,F70,EFL79}.

The present work was conceived as an approach to this kind of experimental works. Our laboratory set-up is
directly inspired in the work by Durian and co-workers, who analyzed the properties of Brownian motion in a macroscopic particle \cite{OLDLD04}; a ping pong ball fluidized by turbulent air
wakes. These wakes are produced at the Von K\'arm\'an streets due to air flow past a spherical
particle \cite{vD82}. We built a very similar set-up, this time using $\sim 10^2$ particles in most of our experiments. The final aim  of our series of measurements is to search for eventual phase transitions, in analogy with the observations in thin vibrated layers \cite{OU98}.


\section{Description of the system}

We perform experiments with a variable number $N$ of identical spherical particles. Specifically, our particles are ping-pong balls  with diameter  $\sigma = 4~\mathrm{cm}$ (ARTENGO$^{\copyright}$ brand balls, made of ABS
plastic, this material having a mass density $\rho\simeq 0.08~\mathrm{g\,cm}^{-3}$). Particles rest on a metallic mesh (circular holes of $3~\mathrm{mm}$ diameter arranged in a triangular lattice) and are enclosed by a circular wall made of
polylactic acid (PLA). The diameter of this circular boundary is
$D=72.5~\mathrm{cm}$ and its height is $h\simeq 4.5~\mathrm{cm}>\sigma$. Thus, the
total area of the system available to the spheres is $A=0.413~\mathrm{
  m}^{2}= 328.65\times\pi(\sigma/2)^2$, which means that up to $N_\mathrm{max}=0.9069\times 328.65\simeq298$ balls can fit in our system, if we ignore boundary effects (which would reduce this number). The grid is mounted and carefully levelled so that it remains as horizontal as
possible (i.e.; perpendicular to gravity). In this way, we expect to achieve
dynamics in which gravity does not create any anisotropy in the system behaviour.

\begin{figure*}[!t]
    \centering
    \includegraphics[height=3.7cm]{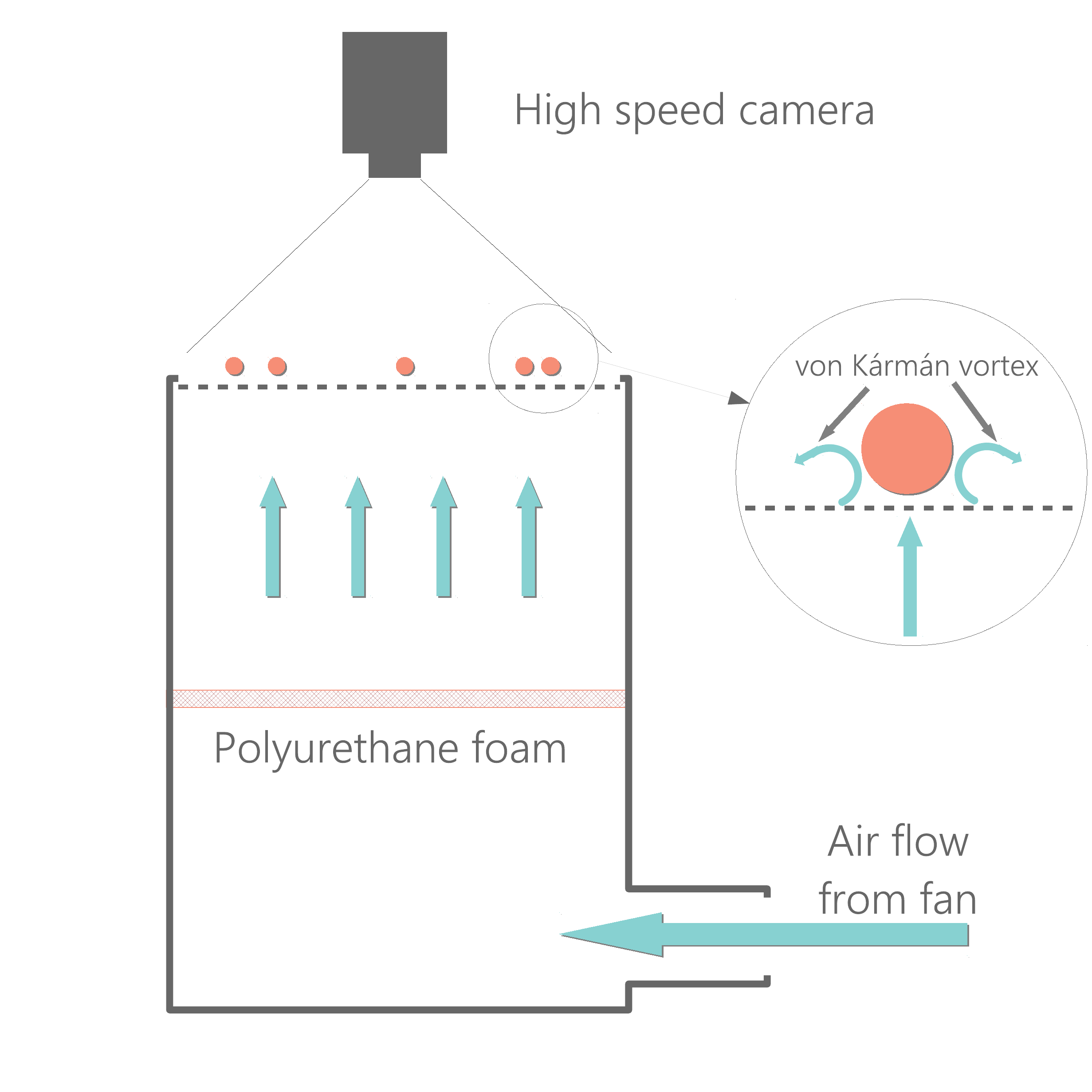}
    \caption{Sketch of the experimental set-up} 
    \label{scheme}
\end{figure*} 

Steady dynamics of the particulate system is achieved by means of a vertical air
current coming from below, as depicted in Figure \ref{scheme}. We use for this
a fan, model SODECA$^{\copyright}$ HCT-71-6T-0.75/PL, that is able to produce air current intensities $u_{air}$ passing
through the metallic grid plane with rates in the range of $[$2 - 5.5$]$ $\mathrm{m/s}$. An intermediate foam ($\sim 2~\mathrm{cm}$ thick) homogenizes the air current from the fan. The air flow 
distribution over the grid was measured using a turbine digital
anemometer, plugged to a computer to collect data. We observed that the air current flow suffers local deviations over the grid of less than $10\%$, with respect to the average $u_{air}$.

The air flow coming from the fan produces turbulent wakes past the
spheres \cite{vD82}. 
Air current intensity is adjusted so that particles never levitate, and remain in contact with the metallic grid at all times. With this, we achieve a particle dynamics that is effectively two-dimensional, since the relevant particle motion is contained within the grid plane.

%

Summarizing, we built a set-up that has the following properties: 1) It is a
many-particle system; 2) the energy input is
homogeneous; 3) the dynamics is contained in a horizontal plane (the grid) and as a
consequence gravity does not introduce a predominant direction; 4) symmetry-break is observed in the form of a hexagonal arrangement of the particles. 

A series of experiments has been carried out, by modifying the values of air flow
intensity ($2~\mathrm{m/s}\leq u_{air} \leq 5.5~\mathrm{m/s}$) and packing fraction $\varphi\equiv N(\sigma / D)^{2}$, ($0.03\leq \varphi \leq 0.79$).

\section{Particle tracking}

A Phantom VEO 410L high speed camera has been used \cite{phantom}. This model is capable of
recording at 5200 frames per second at its maximum resolution ($1280\times800
~\mathrm{pixels}$). At this resolution, we have recorded for each
experiment a $99.92~\mathrm{s}$ clip at $250~\mathrm{frames/s}$ (well below the
maximum operational speed of our camera).
For low particle densities ($\varphi<0.20$) 
three clips have been ta ken for each set of parameters ( which improves the statistics of the data). Before each take the system hasbeen thermalized for a few minutes, in order to assure steady state conditions (transient regime is very short for granular gas systems in most cases \cite{VSK14}).

We have developed a custom detection algorithm, where we combine the use of the open source
library OpenCV \cite{opencv} for particle detection together with TrackPy \cite{dan_allan_2019_3492186} (a python version of the Crocker and Grier algorithm \cite{CG96}) for linking particle
positions.
Only a central region of interest (ROI) has been used for our subsequent analysis.
This is done minimize eventual boundary effects
\cite{Lanoiselee2018,Abate2006}. Our ROI is rectangular and has dimensions
$L_{x}=9.6~\sigma$, $L_{y}=7.7~\sigma$.

\subsection{Error estimation}
Static localization error is defined as the standard deviation from the position of a
motionless particle \cite{Berglund2010,Ober2004,Thompson2002}. In order to quantify
this static error, we recorded five videos with a single static sphere (fan switched off), positioned at different
points of the system in each one. Measuring the deviation from the mean position, we
have found this static error to be $\epsilon_{s} = 0.059$ pixels or, in units of the
particle diameter, $\epsilon_{s} = 7.6 \times 10^{-4}~\sigma$.   
Additionally, there are other factors that can have an effect on the
quality of measurements, such as motion blur --also called dynamic error--
\cite{Savin2005}; this is due to the fact that cameras take a certain amount of time
$\Delta t$ to acquire each frame. In our particular case we used $\Delta t = 1.5\times
10^{-3}$ s. Therefore, the positions that we obtain are not instantaneous, but
averaged over that short period of time. This error has implications, for example, in
determining the value of the diffusion coefficient $D$, and introducing uncertainty in
the mean squared displacement (MSD) measurement \cite{Berglund2010}. Since we do not
discuss here dynamical properties with characteristic times shorter than our $\Delta
t$, this dynamic error can be neglected in the present work. 

\begin{figure*}[!t]
    \centering
    \includegraphics[width=0.31\textwidth]{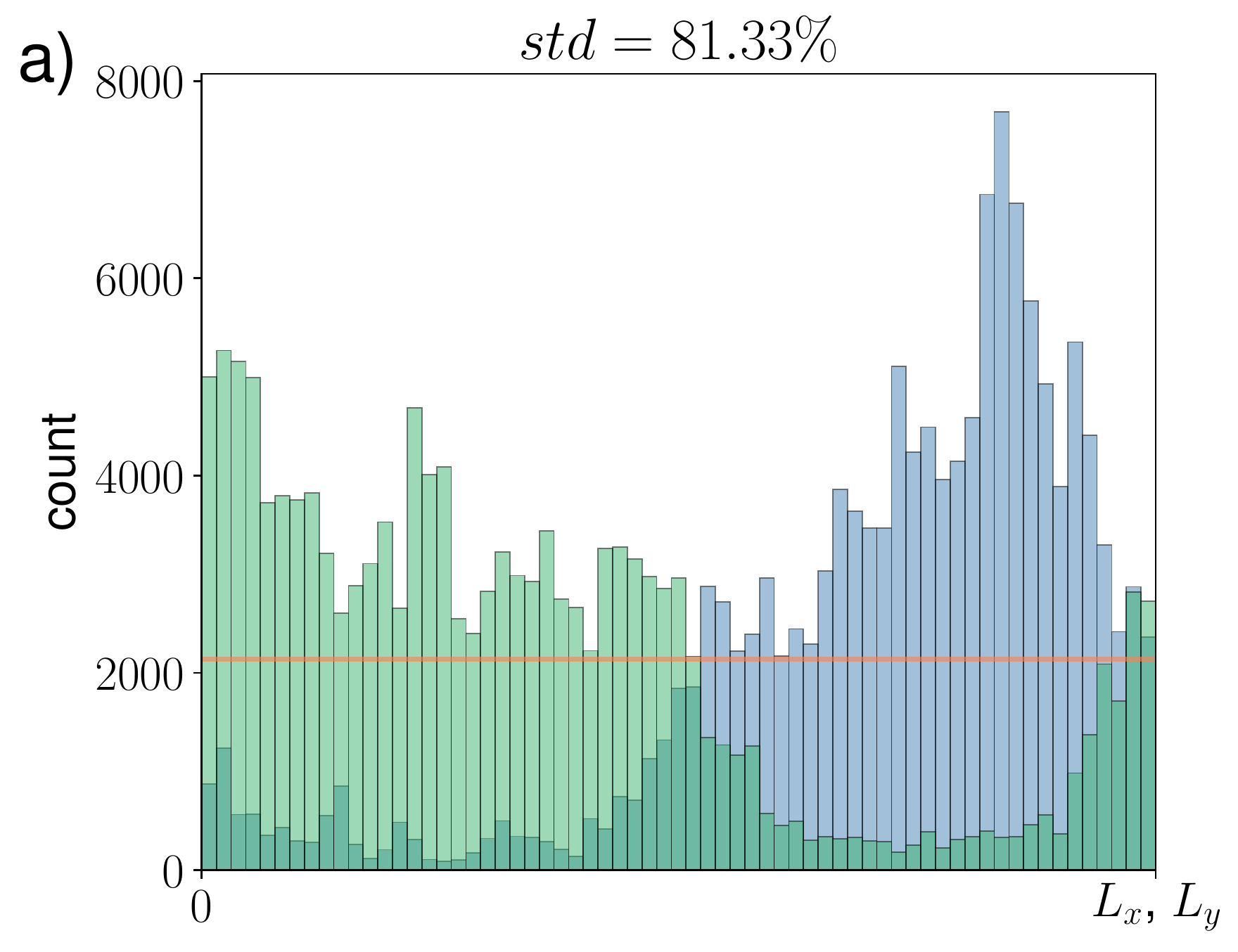}
    \includegraphics[width=0.31\textwidth]{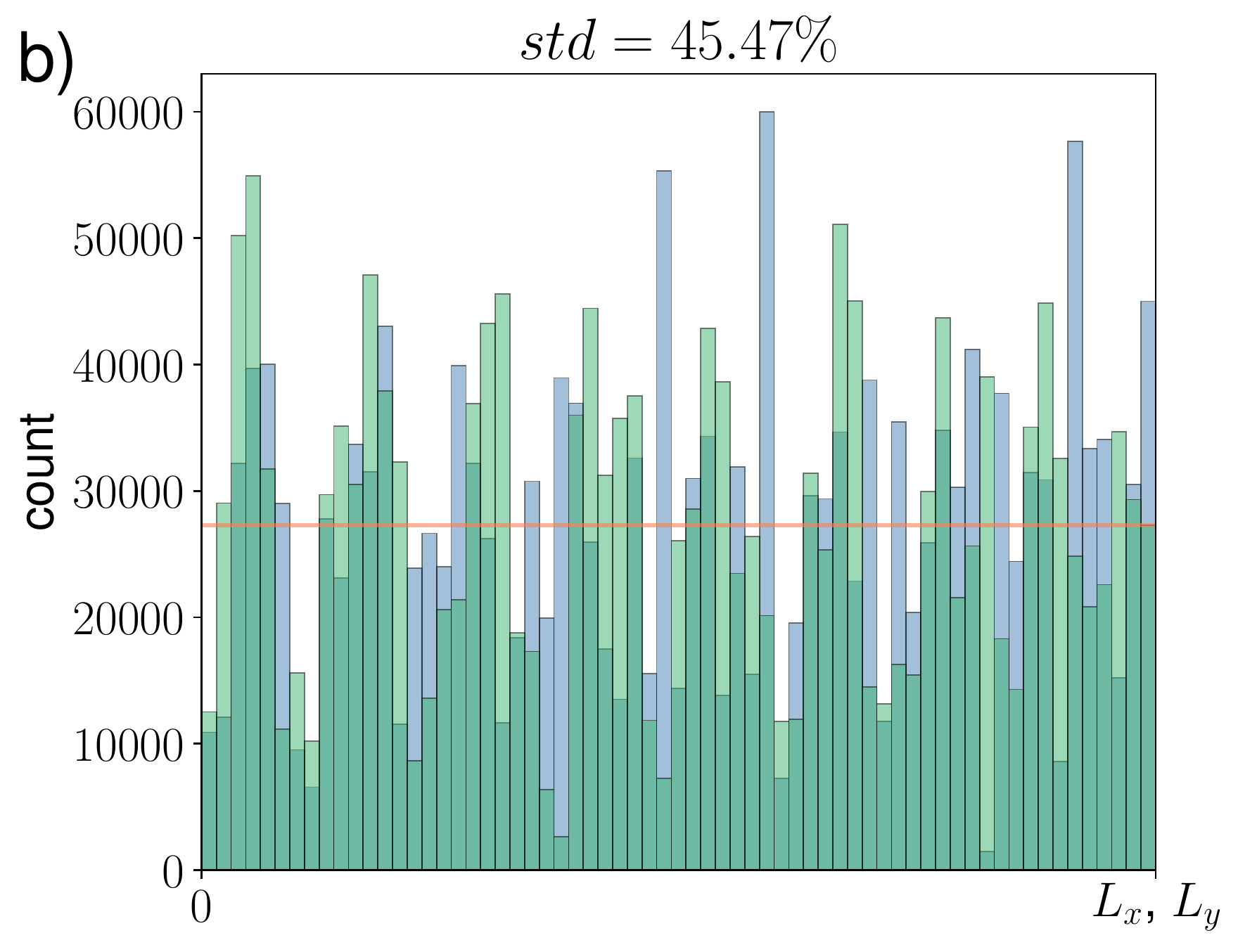} 
    \includegraphics[width=0.31\textwidth]{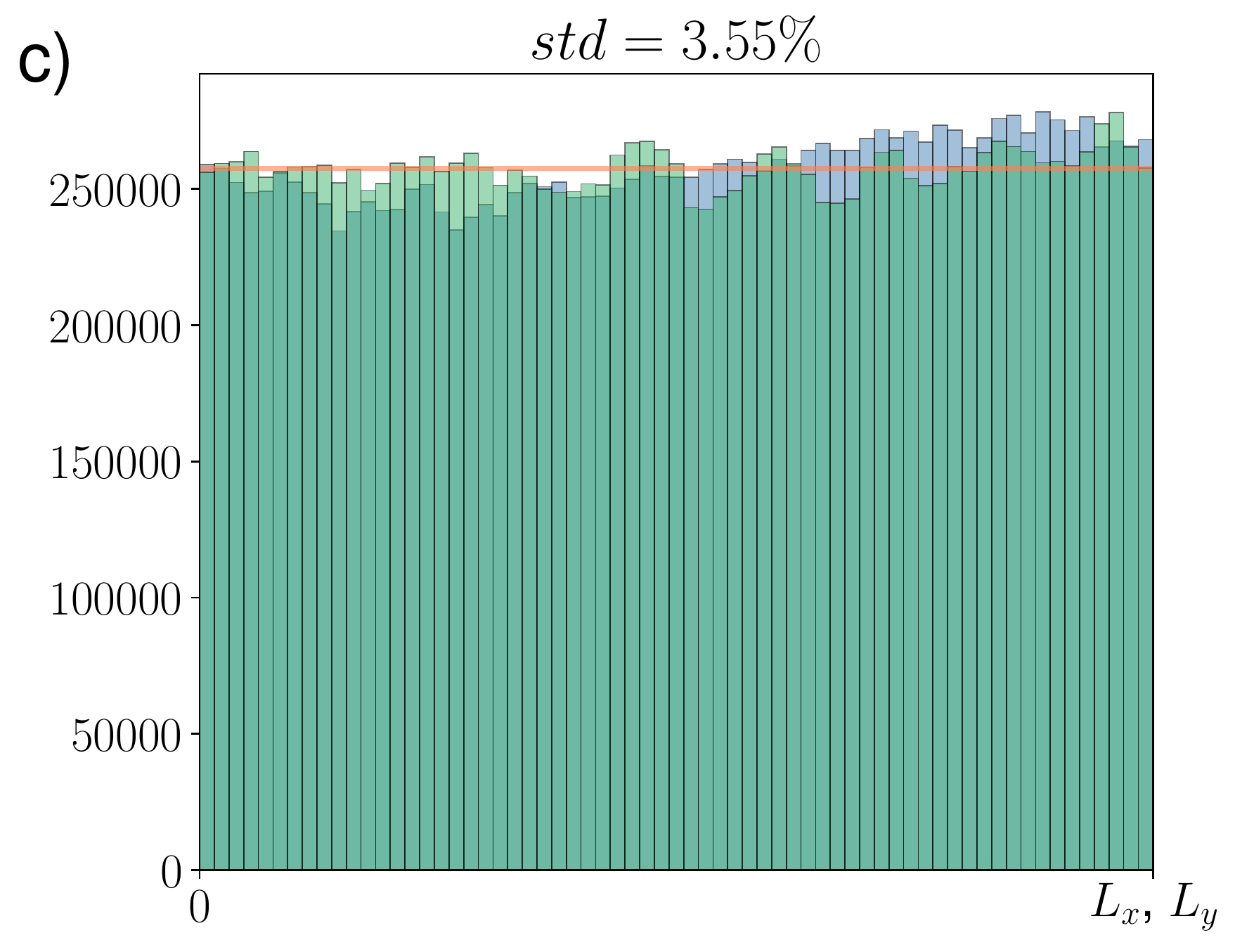} 
    \caption{Histogram of particle locations in the $x$ (blue) and $y$ (green) directions, during full-length clips for: (a) a
      system with a low packing fraction ($\varphi = 0.03$, $u_{air}=4.63$ m/s), here particles remain
      near their initial positions; (b) a case in which packing fraction approaches a crystallization regime ($\varphi =
      0.69$, $u_{air}=2.80$ m/s), the reason behind the histogram maxima is that particles organize in a lattice
      structure. And (c), combined data for all the experiments
      recorded at intermediate densities, for which an isotropic liquid state is found ($0.20 \leq \varphi \leq 0.54$).}  
    \label{isotropy}
\end{figure*}

\subsection{Isotropy and homogeneity conditions}

In order to analyze the degree of isotropy in the collective dynamics, we have looked
at the position distribution for $x$ and $y$ coordinates. Results are plotted
as histograms in Figure~\ref{isotropy}. We observed that except for very low (Figure \ref{isotropy}a) or very high particle densities (Figure \ref{isotropy}b), 
particles are evenly allocated throughout the surface, as seen in Figure \ref{isotropy}c, with a standard deviation never greater than $5\%$ \cite{Scholz2017}. 
In the case of high packing fractions, evidences of
crystallization show up in the form of sharp and evenly distributed peaks (Figure \ref{isotropy}b). On the other hand, at very low densities (Figure \ref{isotropy}a), granular temperature is so low that
the particles rarely move far away from their initial positions, thus skewing the
distribution towards those regions.

\begin{figure*}[!t]
    \centering
    \includegraphics[height=4.2cm]{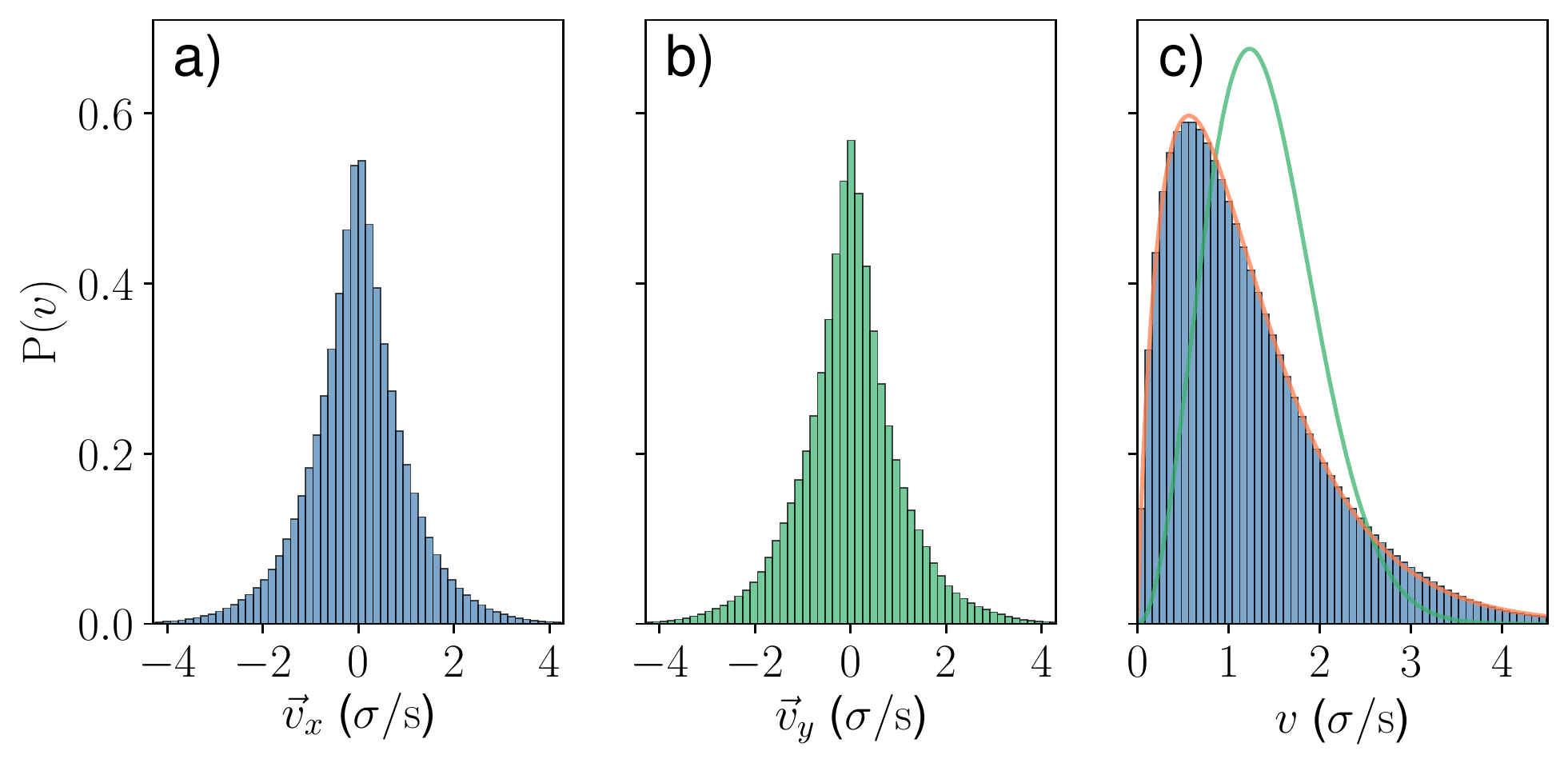}
    \caption{Probability distribution function for the velocity of particles in the
      whole series of experiments. Displacements in the (a) $x$ direction, (b) $y$
      direction, and (c) modulus of the velocity; for reference, gamma (orange) 
      and Maxwellian (green) distribution fits are represented.} 
    \label{vel_distr}
\end{figure*}

With respect to the velocity distributions, and contrary to what happens with positions, we did not
detect different behaviours depending on the packing fraction. Therefore, we represent in Figures
\ref{vel_distr}a and \ref{vel_distr}b $x$
and $y$ velocity distribution functions respectively, combining the data from all of the
clips altogether (we analyze in this way the \textit{global isotropy of the set-up} rather than the
isotropy of a particular experiment). We have observed that distribution functions in $v_{x}$ and $v_{y}$
are not Gaussian, which agrees with previous observations in granular dynamics experiments
\cite{OU99}. In any case, our velocity distributions are rather well centered
at zero (mean values $\mu_x=4.32 \times 10^{-3} \sigma / \mathrm{s}$ and $\mu_y=8.56 \times
10^{-3} \sigma / \mathrm{s}$) and both have a similar standard deviation ($\sigma_x=1.06~
\sigma / \mathrm{s}$ and $\sigma_y=1.07~ \sigma / \mathrm{s}$), which is an indication of a high degree of horizontality. In addition, we represent
in Figure \ref{vel_distr}c the particle speed modulus global distribution function $v = (v_{x}^2 + v_{y}^2)^{1/2}$,
with mean value $1.22~ \sigma / \mathrm{s}$ and standard deviation $0.88~\sigma / \mathrm{s}$. 
It is interesting to notice that a Gamma probability distribution ($f(x) = x^{a-1}e^{-x}/\Gamma(a)$ with shape parameter $a=1.863$, in this case) provides a much better
fit than a Maxwell-Boltzmann one.

In Figure~\ref{isotropy} velocities are deduced from to frame-to-frame displacements. Let us analyze now the
distribution of displacements at longer times. For this, we
have determined the distribution function for 200-frames displacements, $\Vec{d}_{200}
= \Vec{r}(t+200) - \Vec{r}(t)$, over the entire length of each experiment (of about
$10^2~\mathrm{s}$); i.e., we have of the order of $2.5\times 10^4$ events per particle. Would anisotropy occur, it should show up in a polar coordinate
representation (as in Figure~\ref{windrose}b) in the form of a well defined maximum in the preferential displacement
direction.  We have selected a 200 frames window (this is roughly the
time at which, at intermediate densities, particles 'forget' in our set-up their initial velocity, as calculated from the velocity autocorrelation function \cite{LCetal19}). Figure \ref{windrose}b reveals a high
degree of displacements isotropy. Figure \ref{windrose}a displays the histogram of particle positions for the same experiment. Interestingly, this representation reveals very clearly a hexagonal crystal structure.

\section{Discussion}

We have studied in this work a system of ping-pong balls thermalized by means of turbulent air wakes. The air current is adjusted so that balls are in contact with the table at all times (i.e., their movement is primarily two-dimensional). In the first part of this work, we have analyzed the accuracy of our experimental methods for particle tracking. As in previous works by other authors \cite{Berglund2010,Ober2004,Thompson2002}, we have made distinction between inherent static and dynamic errors, finding that in both cases they are very small. Homogeneity of particle density indicates that we achieved a good degree of horizontality in of our set-up (this work is intended to study only inhomogeneity-free dynamics).
For intermediate densities, the degree of isotropy is very high (Figures
\ref{isotropy}c and \ref{vel_distr}). It is very interesting to remark that the
distribution function of $v$ (averaged over all of the series) fits a gamma
distribution \cite{statistics}. 

\begin{figure*}[!t]
    \centering
    \includegraphics[height=4.25cm]{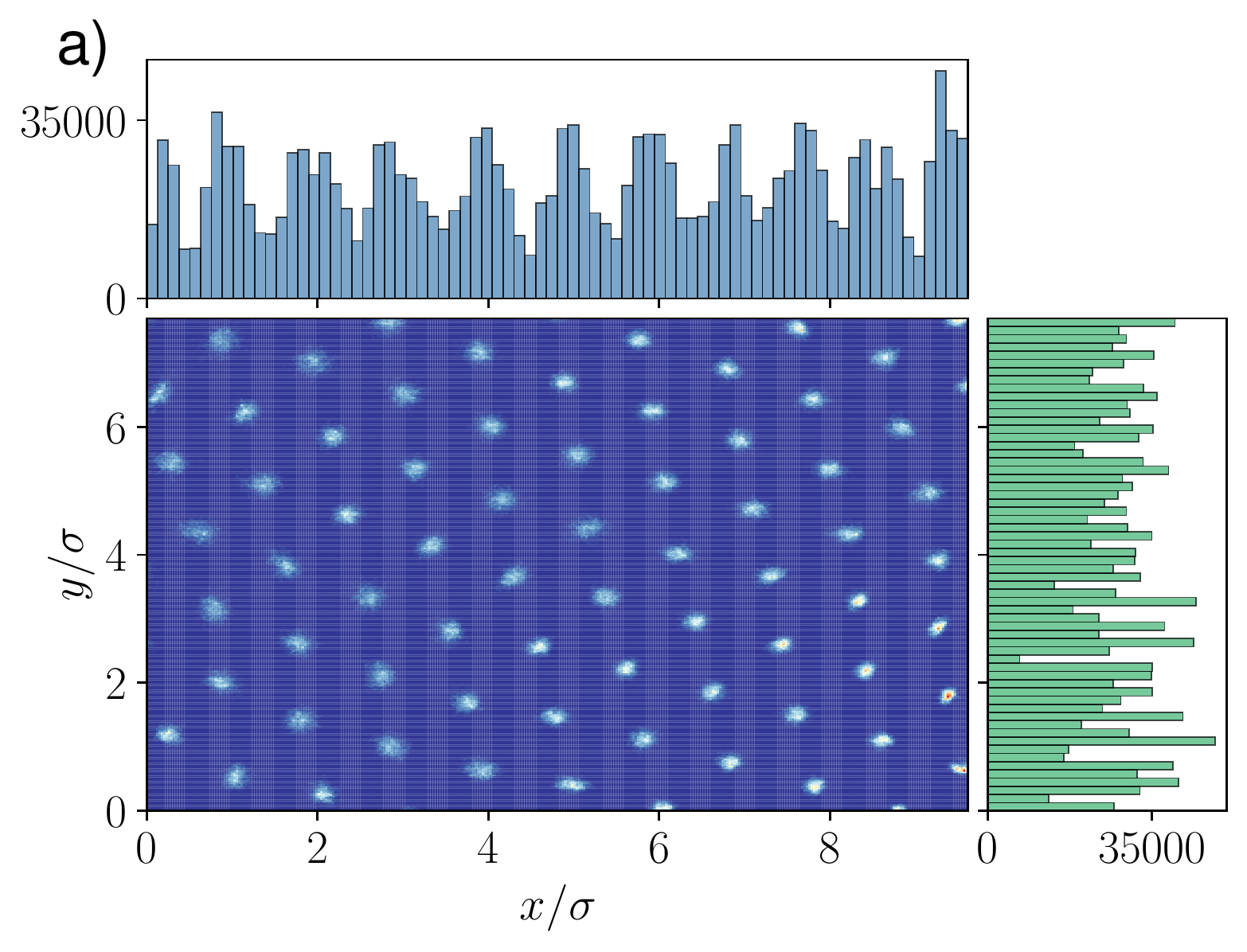} \quad
    \includegraphics[height=4.25cm]{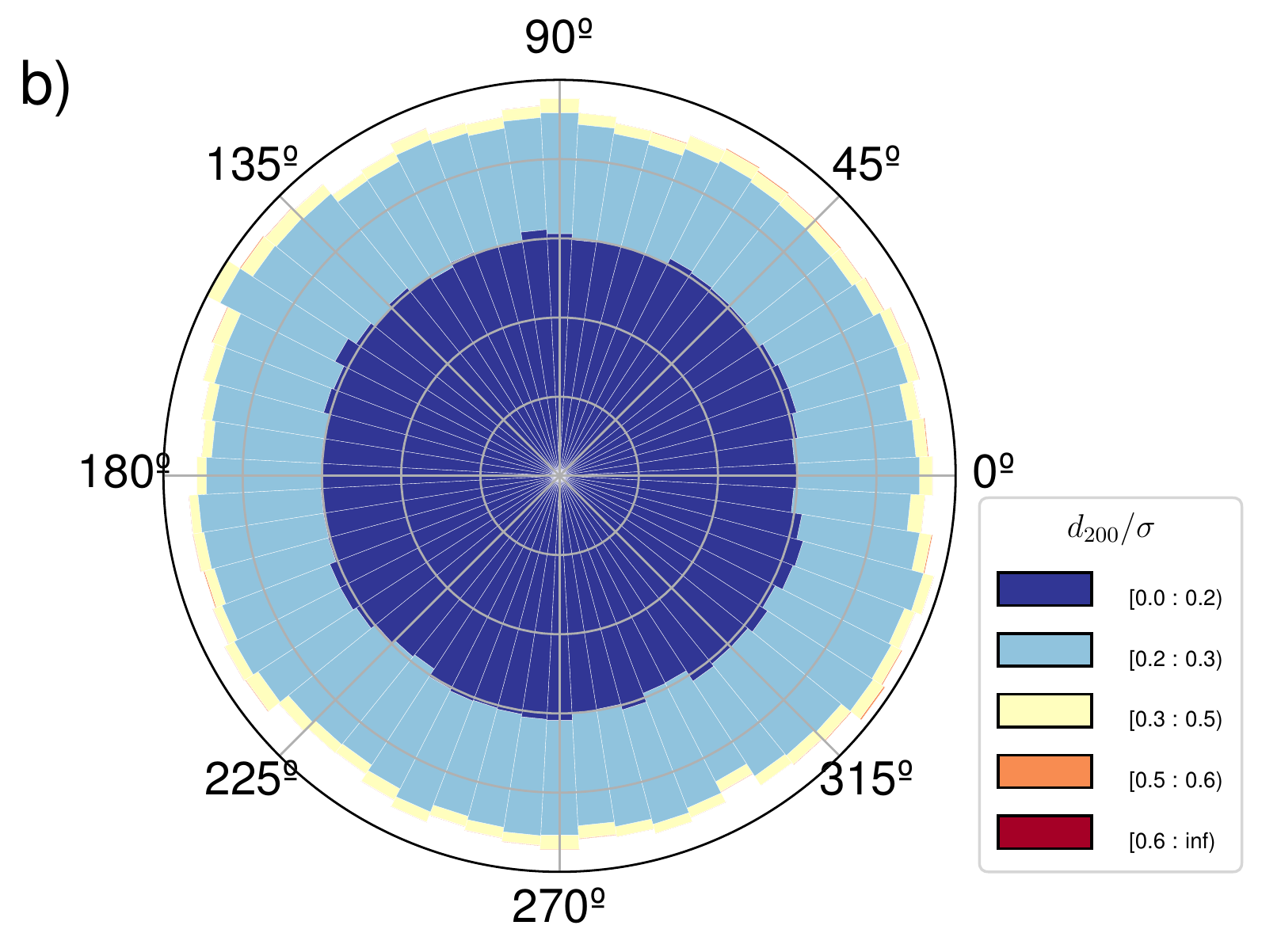}
    \caption{Experiment with packing fraction $\varphi = 0.67$ and air flow velocity $u = 2.80$ m/s. (a) Particle positions histograms (upper and left panels) and plot of particle positions superimposed for all frames (central panel). A crystalline structure is clearly visible. (b) Stacked histogram in polar coordinates of 200 frame displacements ($\Vec{d}_{200}$) for that same system; different colours represent different displacement lengths (in units of the particle diameter). The histogram strongly suggests that particle dynamics is highly isotropic. } 
    \label{windrose}
\end{figure*}

At high densities, it is interesting to notice the remnants of what
seems to be a hexagonal crystal structure (Figures \ref{isotropy}b and \ref{windrose}a). Phase transitions that can eventually appear in this system will be studied in more detail
in an impending work.
  
\section*{\small{Acknowledgments}}
 The authors thank J. S. Urbach, E. Abad and S. B. Yuste for fruitful discussions. We acknowledge funding from the Government of Spain through project No. FIS2016-76359-P and from
 the regional Extremadura Government through projects No. GR18079 \& IB16087, both partially
 funded by the ERDF.


\bibliography{ppp}
\bibliographystyle{spmpsci}

\end{document}